\documentclass{iopart}

\usepackage{graphicx}
\usepackage{dcolumn}
\usepackage[latin1]{inputenc}
\usepackage{color}
\usepackage{enumerate}

\begin{document}

\title{Generation of degree-correlated networks using copulas}

\author{Mathias Raschke$^1$, Markus Schl\"apfer$^2$ and Konstantinos Trantopoulos$^3$}
\address{$^1$ Raschke CAT-Softwareentwicklung, Stolze-Schrey-Strasse 1, 65195, Wiesbaden, Germany}
\address{$^2$ Senseable City Laboratory, Massachusetts Institute of Technology, \mbox{77 Massachusetts Avenue}, Cambridge, MA 02139, USA}
\address{$^3$ Department of Management, Technology, and Economics, ETH Zurich, \mbox{Weinbergstrasse 56/58}, 8092, Zurich, Switzerland}

\ead{mathiasraschke@t-online.de, schlmark@mit.edu, trantoko@ethz.ch}

\begin{abstract}
Dynamical processes on complex networks such as information propagation, innovation diffusion, cascading failures or epidemic spreading are highly affected by their underlying topologies as characterized by, for instance, degree-degree correlations. Here, we introduce the concept of copulas in order to artificially generate random networks with an arbitrary degree distribution and a rich \textit{a priori} degree-degree correlation (or `association') structure. The accuracy of the proposed formalism and corresponding algorithm is numerically confirmed. The derived network ensembles can be systematically deployed as proper null models, in order to unfold the complex interplay between the topology of real networks and the dynamics on top of them. 
\end{abstract}
\maketitle
\section{Introduction}
\label{sec:intro}
Drawing on the pertinent literature, network studies have provided substantial insights into the skeletal morphology of various systems, with examples as diverse as the human brain, online social communities, financial networks or electric power grids \cite{Boccaletti:2006,Dorogovtsev:2008, Schweitzer:2009}. Going beyond characterizing the network topology by the essential degree distribution, extensive research has focused on the degree-degree association\footnote[1]{The term `association' is used in this paper as it refers to the general relation between two random variables, while the term `correlation' is restricted to a single measure.
} \cite{Newman:2002}. A positive degree-degree association represents the tendency of nodes with a similarly small or large degree to be connected to each other. A negative degree-degree association accordingly implies that the nodes tend to be connected to nodes with a considerably different degree. Interestingly,  a positive association is typically found in social networks, while a negative association can often be observed in biological and technical ones \cite{Newman:2003}. 

Generating artificial random networks with an \textit{a priori} association structure is a prerequisite for systematically investigating real networks. Such null models can eventually be used to shed light on the interplay between dynamical phenomena on networks and the underlying topology. Vivid examples range from information diffusion \cite{Karsai:2010} and epidemic spreading \cite{Schlaepfer:2012} in social networks to cascading failures in power grids \cite{Schlaepfer:2008, Schlaepfer:2010}. The reshuffling method according to \cite{Xulvi-Brunet:2004,Menche:2010} is commonly used in order to impose a desired level of degree-degree association on random networks, as quantified by a single \textit{association measure}. While this is a straightforward algorithm, it appears to be incapable to fully control the overall \textit{association structure}. This is a substantial drawback, as two networks with an equal association measure can exhibit significantly different association structures, eventually implying different impacts on the dynamics on top of them. A first step towards this direction has already been proposed in \cite{Weber:2007}, by drawing upon two-point correlations of empirical networks. Furthermore, the Gaussian copula function has recently been deployed for the particular case of generating random networks with Poissonian degree distribution and given association measures \cite{Gleeson:2008}.

Here, we propose a general method for constructing random network ensembles with an arbitrary degree distribution and desired degree-degree association structure by using various copula functions. This allows to provide more comprehensive null models with a complete description of their degree-degree association. The paper is organized as follows. Section \ref{sec:matrix} introduces the construction of probability matrices with an imposed degree-degree association, based on copulas. A general formalism for the realization of random networks based on a given probability matrix is provided in Section \ref{sec:random}, together with a description of the corresponding algorithm and its numerical evaluation. Section \ref{sec:conclusions} concludes.
\section{Constructing the probability matrix}
\label{sec:matrix}
\begin{figure}[t!]
\begin{center}
\includegraphics[scale=0.4]{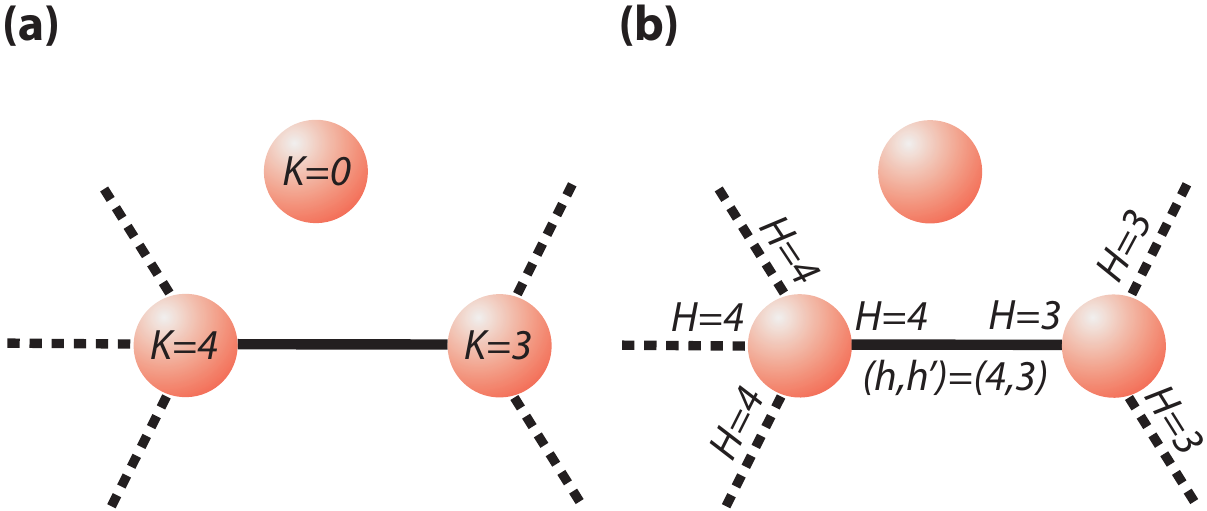}    
\caption{\label{fig:Figh}Different assignment of the number of edges. (a) $K$ edges per node, (b) $H$ edges connected to the end of an edge.}
\end{center}
\end{figure}

The probability matrix as introduced in \cite{Raschke:2010} approximates the degree-degree association structure by a bivariate distribution of discrete random variables. This allows to generate different realizations of networks with the same underlying association structure. The probability matrix $P(h,h')$ is the joint distribution of the number of edges $h$ connected to the end of an edge, including the considered edge itself. The assignment and its difference to the node degree $k$ (number of edges incident on a node) is illustrated in \mbox{Fig. \ref{fig:Figh}}.
The marginal distribution of $P(h,h')$ is the distribution $P_h(h)$, which is related to the distribution of the node degree $P_k(k)$ by $P_h(h)=P_k(h)h/\left\langle k \right\rangle$, with $\left\langle k \right\rangle$ being the average degree. Note that $|k|=|h|$ for a specific node, implying $h_{max}=k_{max}$. A straightforward way to construct the probability matrix is the application of a bivariate discrete random distribution \cite{Raschke:2010}. However, this approach suffers from the limited number of discrete and especially heavy-tailed bivariate distributions. During the recent decades, the use of copulas has hereby proved to be powerful to overcome the same shortcoming in the continuous case \cite{Sklar:1959,Mari:2001,Embrechts:2005,Nelsen:2006}. The basic idea is to separate the marginal distributions from the association structure. 

Based on Sklar's Theorem \cite{Sklar:1959} the copula $C(u,v)$ for the continuous random variables $X$ and $Y$ is defined by the bivariate cumulative distribution function (CDF) $F_{xy}(x,y)$, with the marginal distributions $F_{x}(x)$ and $F_{y}(y)$ 
\begin{equation}
\label{eq:1}
C(u,v)=F_{xy}(F^{-1}_{x}(u),F^{-1}_{y}(v)),
\end{equation}
where $F^{-1}$ is the inverse function and $u=F_x(x)$ and $v=F_y(y)$. The simplest version of a copula is the application of the structure $F_{xy}(x,y)$ to the random variables $W$ and $Z$,
\begin{equation}
\label{eq:2}
C(F_w(w),F_z(z))=F_{xy}(F^{-1}_{x}(F_w(w)),F^{-1}_{y}(F_z(z))).
\end{equation}
The probability $P$ that the random variables $U$ and $V$ are found in the intervals $[u_1,u_2)$ and $[v_1,v_2)$, respectively, is
\begin{eqnarray} \label{eq:3}
P(u_1 \leq U < u_2, v_1 \leq V < v_2) 
=  C(u_1,v_1) {} \nonumber\\ +C(u_2,v_2) -C(u_1,v_2)-C(u_2,v_1).
\end{eqnarray}
This formalism is used to construct the probability matrix $P(h,h')$ with the marginal distribution $P_h(h)$, whereas three different procedures can be followed. In \mbox{\textit{procedure I}}, $P_h(h)$ is always defined by a left bounded continuous CDF $F_x(x)$ (or $F_y(y)$ respectively).  
\begin{equation}
\label{eq:4}
P_h(h)=F_x(h)-F_x(h-1),
\end{equation}
where $0 \leq h_{min}\leq h \leq h_{max}$ and $x_{min}:=h_{min}-1$. Choosing a specific copula function and combining Eqs. \ref{eq:3} and \ref{eq:4} gives the probability matrix $P(h,h')$
\begin{eqnarray} \label{eq:5}
P(h,h')=F_{xy}(h,h')+F_{xy}(h-1,h'-1)  {} \nonumber\\
-F_{xy}(h-1,h')-F_{xy}(h,h'-1).
\end{eqnarray}
In the case that $h$ is left and right bounded with $h_{max}< \infty$, the probability matrix becomes truncated and has to be normalized, i.e., $\sum_{h,h'} P(h,h')=1$, and the marginal distribution is recalculated with
\begin{equation}
\label{eq:6}
 P_h(h)=\sum_{h'=h_{min}}^{h_{\max}}P(h,h').
\end{equation}

For \textit{procedure II}, the marginal distribution $P_h(h)$ is given, and the probability matrix is written as
\begin{eqnarray} \label{eq:7}
P(h,h')&=&C(G_h(h),G_h(h')) {} \nonumber\\
&&+C(G_h(h-1),G_h(h'-1))  {} \nonumber\\
&&-C(G_h(h-1),G_h(h'))  {} \nonumber\\
&&-C(G_h(h),G_h(h'-1)),
\end{eqnarray}
where $G(h)=\sum_{j=h_{min}}^{h}P_h(j)$. The matrix $P(h,h')$ can again be truncated at $h_{max}$ as in procedure I. Note that in the case of heavy-tailed $F_x$, the resulting marginal distributions $P_h(h)$ in procedures I and II are not strictly heavy-tailed due to the truncation.

For \textit{procedure III}, the distribution $P_h(h)$ is obliged to be truncated at $h_{max}$, implying $G(h_{max})=1$. The range of the copula is now limited, whereas the numerical differences between the resulting probability matrix and $P(h,h')$ derived by procedure II become smaller with increasing $h_{max}$.

\begin{figure}[b!]
\begin{center}
\includegraphics[scale=0.5]{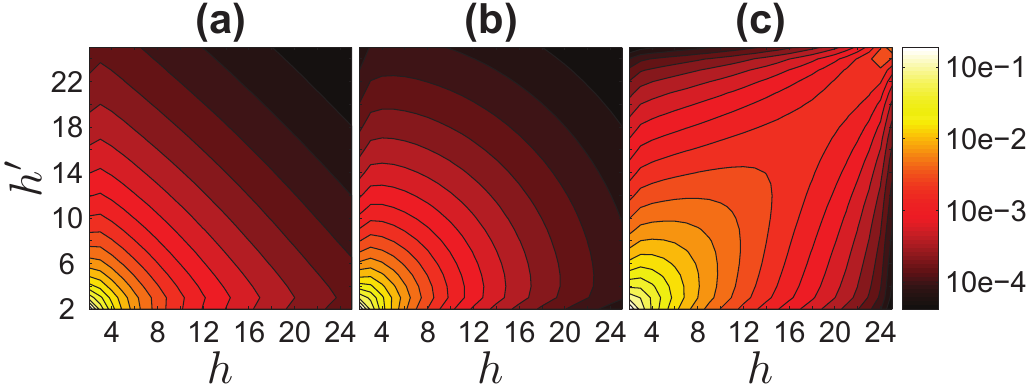}    
\caption{\label{fig:1}Resulting probability matrices $P(h,h')$ based on the Gumbel copula with $\lambda=2$. (a) Procedure I with a continuous Pareto marginal distribution, $F_x(h)=1-(h-1)^{- \gamma}$ with $\gamma=0.7$. (b)
Procedure II with a Zipf marginal distribution, $P(h)=h^{-\gamma}/A$, where $A=\sum_{h_{min}}^{\infty}h^{-\gamma}$ and $\gamma=2$.  (c) Procedure III with a truncated Zipf marginal distribution, $P(h)=h^{-\gamma}/A$, where $A=\sum_{h_{min}}^{h_{max}}h^{-\gamma}$ and $\gamma=2$. In all three cases $h_{min}=2$ and $h_{max}=25$. The color bar corresponds to all panels.}
\end{center}
\end{figure}

An example for the three procedures is the application of a Gumbel copula \cite{Mari:2001} with copula parameter $\lambda$,
\begin{equation}
\label{eq:8}
C(u,v)=exp(-((-ln(u))^\lambda+(-ln(v))^\lambda)(1/\lambda)).
\end{equation}
Figure \ref{fig:1} depicts the probability matrices with $\lambda=2$ as derived by the three procedures. Interestingly, the different procedures are leading to considerably different association structures, although the same copula function and similar marginal distributions are applied.

\begin{figure}
\begin{center}
\includegraphics[scale=0.55]{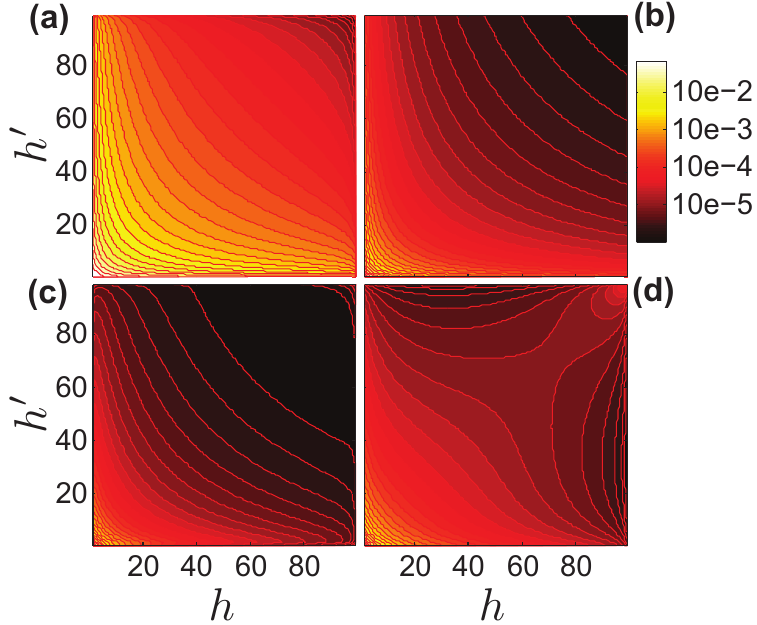}    
\caption{\label{fig:2}Comparison of probability matrices $P(h,h')$ with a truncated Zipf marginal distribution $P(h)=h^{-\gamma}/ \sum_{h_{min}}^{h_{max}} h^{-\gamma} $ for different values of Kendall-Gibbons' $\tau_b$. (a) Gaussian copula with $\tau_b$=-0.3, (b) Gaussian copula with $\tau_b$=0, (c) Gaussian copula with $\tau_b$=0.3 and (d) Gumbel copula with $\tau_b$=0.3. In all cases $\gamma=1.5$, $h_{min}=2$ and $h_{max}=100$. The matrices have been constructed following procedure III. The color bar corresponds to all panels.}
\end{center}
\end{figure}

Copulas are related to association measures such as Kendall-Gibbons' $\tau_b$ \cite{Mari:2001}, whereas different types of copulas (i.e., different association structures) may imply the same value of the respective association measure. The functional relations between the association measures and the parameters of the different (continuous) copulas are given in the literature (e.g., \cite{Balakrishnan:2009}). For large values of $h_{max}$, the discrete probability matrices can be approximated by continuous functions, so that these defined relations are directly applicable in procedures II and III for calculating the copula parameter from the association measure of $P(h,h')$, and vice versa. For small values of $h_{max}$, each specific relation between the association measure of $P(h,h')$ and the copula parameters can be numerically determined. Examples for resulting probability matrices for different values of Kendall-Gibbons' $\tau_b$ based on the Gaussian and Gumbel copulas are shown in Fig. \ref{fig:2}. The effect of the chosen level of association on the structure of $P(h,h')$ is clearly visible [Figs. \ref{fig:2}(a)-\ref{fig:2}(c)], while a different copula function with equal $\tau_b$ leads to considerably different association structures \mbox{[Figs. \ref{fig:2}(c)-\ref{fig:2}(d)].}

Given a real network, the parameters of both the marginal distribution and the copula can be estimated by common methods of statistical inference, such as the maximum likelihood method.

\section{Realization of network ensembles based on $P(h,h')$}%
\label{sec:random}
\subsection{Assignment probability}
\begin{figure}[t!]
\begin{center}
\includegraphics[scale=0.28]{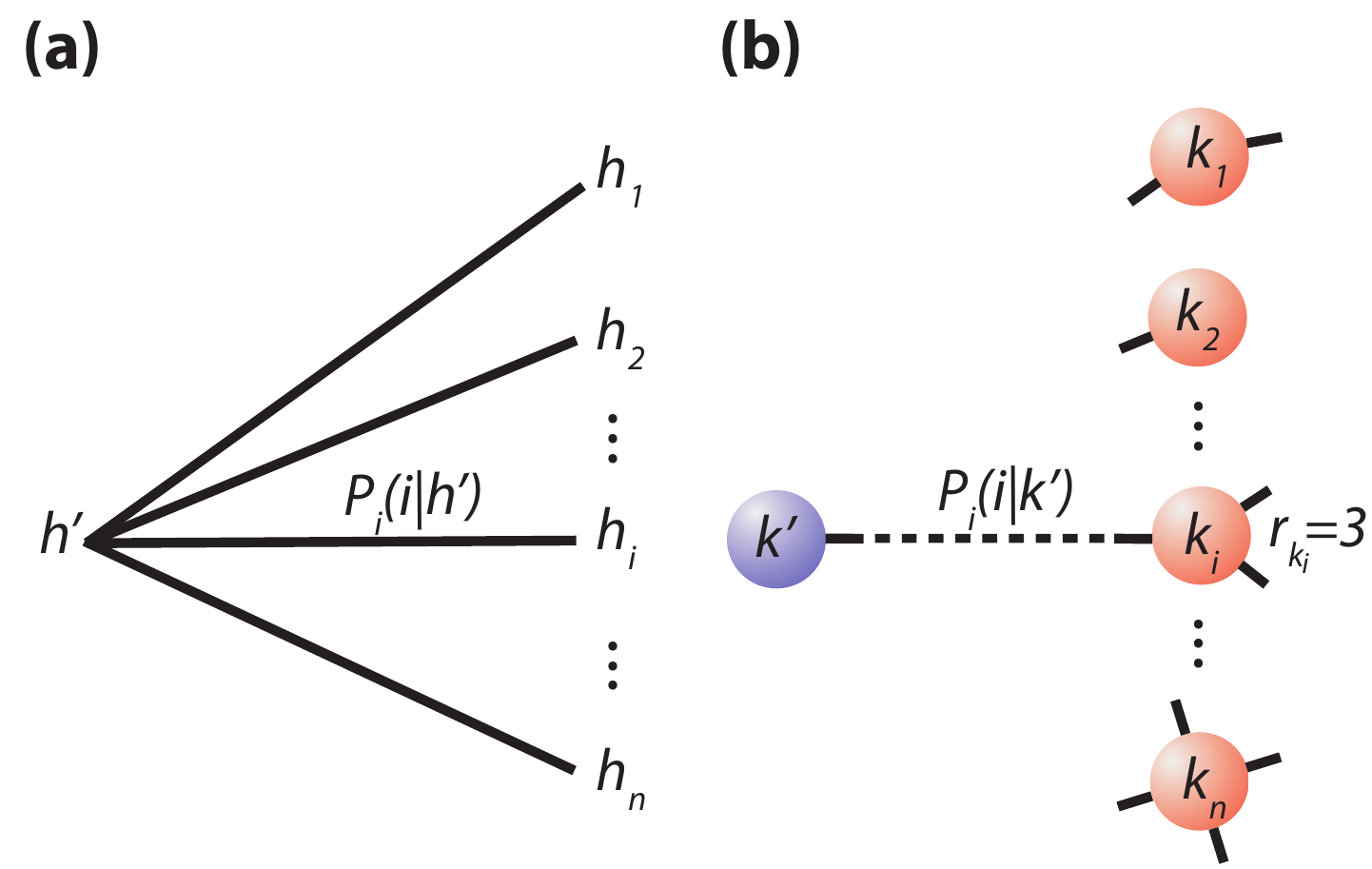}    
\caption{\label{fig:3}(a) Probability to assign realization $h_i$ with position $i$ to realization $h'$. Note that several realizations $h$ may have the same value and may belong to the same node in the network. (b) Assignment probability as used in the network generation algorithm for connecting a node with degree $k_i$ to the selected node with degree $k'$.
}
\end{center}
\end{figure}

Based on a given probability matrix $P(h,h')$ the network generation draws on the assignment probability, as given by the bivariate distribution. We therefore consider an arbitrary sequence $(h_1,h_2,$...$,h_i,$...$,h_n)$ with sample size $n$, where the realizations $h_i$ are randomly distributed according to $P_h(h)$. Letting $n \rightarrow \infty$, the probability to assign a realization $h_i$ with position $i$ to a given realization $h'$, $P_i(i|h')$ [see Fig. \ref{fig:3}(a)], can be derived from the probability matrix by recalling the conditional probability $P(h|h')=P(h,h')/P_h(h')$, and using the relation
\begin{equation} \label{eq:9}
P(h|h')=\sum_{i=1}^{n}P_i(i|h') {\mathbf{1}}_{A_h}(i),
\end{equation}
with the indicator function ${\mathbf{1}}_{A_h}(i)=1$ if $i \in A_h$, and ${\mathbf{1}}_{A_h}(i)=0$ otherwise, where $A_h$ denotes the set of equal realizations $h$. Applying Bayes' Theorem, \mbox{$P(h|h')=P(h'|h)P_h(h)/P_h(h')$}, and by using $\sum_{i=1}^{n} {\mathbf{1}}_{A_h}(i)=nP_h(h)$ one easily computes 
\begin{equation}
\label{eq:11}
P(h'|h)=nP_h(h')P_i(i|h').
\end{equation}
Since $nP_h(h')$ is constant, $P(h'|h)$ is proportional to $P_i(i|h')$. Furthermore, $\sum_i P_i(i|h')=1$.
Thus:
\begin{equation}
\label{eq:12}
P_i(i|h')=P(h'|h_i)/\sum_{j=1}^{n}P(h'|h_j),
\end{equation}
being independent of the sample size $n$.

\subsection{Description of the algorithm}
Based on the assignment probability ([Eq. (\ref{eq:12})], the algorithmic procedure for realizing ensembles of simple networks (i.e., no self-loops and multiple edges) comprises the following steps:
\begin{enumerate}
\item[1.] Random generation of $n$ realizations of $K$, drawn from the probability distribution $P_k(k)$, imposing the constraint that the sum $\sum_j k_j$ must be even. Hence, each node $j$ has a total of $k_j$ ``stubs'' of edges.
\item[2.] Random selection of a node with at least one remaining stub and degree $k'$.
\item[3.] Assignment of the selected node to a node with degree $k_i$, which has again at least one remaining stub and is not yet connected to the selected node. The assignment probability $P_{i}(i|k')$ for connecting these two nodes to one another is given by  
\begin{equation}
\label{eq:15a}
P_{i}(i|k')=r_{k_i} P(k'|k_i)/\sum_{j=1}^{n}r_{k_j} P(k'|k_j).  
\end{equation}
Equation (\ref{eq:15a}) is derived from Eq. (\ref{eq:12}) by substituting the variables $h$ with $k$ and $h'$ with $k'$, respectively, and by considering all the remaining stubs $r_{k_i}$ of the considered node $i$ [see Fig. \ref{fig:3}(b)]. The two selected stubs are connected to form the edge. 
\item[4.] If there are any nodes with remaining stubs go back to Step 2.
\end{enumerate}
In order to generate connected networks represented by a single component (implying $P(1,1)=0$, which introduces intrinsic correlations), step 2 of the algorithm has to be modified in such a way that a node from the already existing network is randomly drawn. If there are any non-connected nodes remaining, but no more free stubs in the existing network available, then an existing edge is chosen randomly (equal weight for each edge) and becomes deleted again. The generation of networks with self-loops and directed or multiple edges is equally well possible by adjusting $r_{k_i}$ and $r_{k_j}$ in step 3 accordingly. The procedure is independent of how the underlying probability matrix $P(h,h')$ has been derived - artificially based on the copula approach, or empirically estimated from real networks. Thereby, the number of edges has to be significantly higher than the maximum degree found in the network, so that the approximation with the probability matrix holds \cite{Raschke:2010}. Note that in contrast to the commonly used algorithm presented in \cite{Weber:2007}, which similarly exploits the concept of bivariate discrete distributions to generate simple networks with arbitrary association structures, the validity of our proposed procedure is directly given by the statistical basics of the assignment probability [Eqs. (\ref{eq:9})-(\ref{eq:12})].

\subsection{Numerical evaluation}
\label{subsec:numerical}
\begin{figure}[t!]
\begin{center}
\includegraphics[scale=0.6]{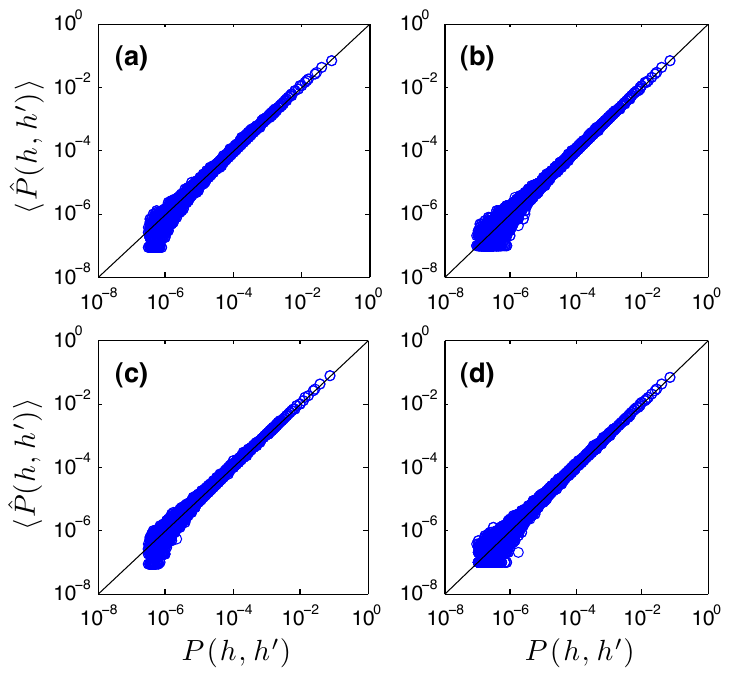}    
\caption{\label{fig:5}Comparison between the underlying probability matrix $P(h,h')$ and the averaged values $\langle \hat{P}(h,h') \rangle$ of the corresponding network ensembles, consisting of 1000 realizations. The networks have a total of $n=5000$ nodes, maximum degree $h_{max}=100$ and minimum degree $h_{min}=1$, respectively. The applied copula is \mbox{$C(u,v)=uv(1-a(1-u)(1-v))$} \cite{Gumbel:1960}, with (a) $a=-0.5$ for positively associated and connected networks, (b) $a=0.5$ for negatively associated and connected networks, (c) $a=-0.5$ for positively associated networks (not necessarily connected), and (d) $a=0.5$ for negatively associated networks (not necessarily connected). Following procedure III as described in Sec. \ref{sec:matrix}, the degree distribution is set to $P(h)=h^{-1.5}/A$ with the normalization factor $A=\sum_{h=h_{min}}^{h_{max}}h^{-1.5}$.  }
\end{center}
\end{figure}

The probability matrix of an artificial or real network can be estimated according to the well-known empirical distribution function,
\begin{equation}
\label{eq:kendall3} 
\hat{P}(h,h')=m(h,h')/2L,  
\end{equation}
wherein $m(h,h')$ is the number of realized pairs $(h,h')$ and $L$ is the number of edges, with each edge contributing to two symmetric pairs. In order to numerically confirm the validity of the proposed aforementioned algorithm, we compare the average $\langle \hat{P}(h,h') \rangle$ of a large number of realized networks with the given determined probability matrix $P(h,h')$. Figure \ref{fig:5} clearly confirms the agreement between them, for the particular case of $P(h) \propto h^{-1.5}$ which roughly corresponds to many real-world networks \cite{Newman:2003}. The possible yet slight deviations in the range of small $P(h,h')$ values can be traced back to the limited number of realized large-degree nodes, naturally restricting the theoretical number of connections between them \cite{Catanzaro:2005}. Hence, high values of $P(h,h')$ for large $h$ and $h'$ may constrain the bivariate approximation, whereas the values of $P(h,h')$ are usually larger for positive association in comparison to negative association.
\vspace{0.5cm}

\section{Conclusions}
\label{sec:conclusions}
In this paper we have introduced a copula-based method enabling the generation of random model networks with an \textit{a priori} desired degree-degree association structure. The copulas are used to construct the underlying probability matrices which, in turn, form the basis for the realization of network ensembles. Our numerical investigations have demonstrated the accuracy of the proposed formalism and its algorithmic implementation. The realized networks can be deployed as proper null models in order to systematically investigate the impact of rich topological structures on various dynamical processes, as found in real networks. Thereby, gaining experience in applying the proposed method will give insights in the most appropriate copula functions to represent empirical networks. 

\ack
M.R. acknowledges ``swisselectric research'' and the Swiss Federal Office of Energy (project No. V155269) for co-funding the present work. K.T. acknowledges partial financial support by the Swiss Federal Office for Civil Protection. We thank Paul Embrechts and Marius Hofert from the RiskLab - ETH Zurich for their valuable comments and discussions on the manuscript.


\section*{References}
\begin{thebibliography}{21}

\bibitem{Boccaletti:2006}    
Boccaletti S, Latora V, Moreno Y, Chavez M and Hwang, D, 2006 \textit{Phys. Rep.} \textbf{424} 175

\bibitem{Dorogovtsev:2008}
Dorogovtsev S N, Goltsev A V and Mendes J F F, 2008 \textit{Rev. Mod. Phys.} \textbf{80} 1275

\bibitem{Schweitzer:2009}
Schweitzer F, Fagiolo, G, Sornette D, Vega-Redondo F, Vespignani A and White D R, 2009 \textit{Science} \textbf{325} 422

\bibitem{Newman:2002}
Newman M E J, 2002 \textit{Phys. Rev. Lett.} \textbf{89} 208701

\bibitem{Newman:2003}
Newman M E J, 2003 \textit{SIAM Rev.} \textbf{45} 167

\bibitem{Karsai:2010}
Karsai M, Kivel\"a M, Pan R K, Kaski K, Kert\'esz J, Barab\'asi A-L  and Saram\"aki J, 2011 \textit{Phys. Rev. E} \textbf{83} 025102

\bibitem{Schlaepfer:2012}
Schl\"apfer M and Buzna L, 2012 \textit{Phys. Rev. E} \textbf{85} 015101(R)

\bibitem{Schlaepfer:2008}
Schl\"apfer M, Dietz S and Kaegi M, 2008 \textit{Proc. Int. Conf. on Infrastructure Systems and Services (Rotterdam)}

\bibitem{Schlaepfer:2010}
Schl\"apfer M  and Trantopoulos K, 2010 \textit{Phys. Rev. E} \textbf{81} 056106

\bibitem{Xulvi-Brunet:2004}
Xulvi-Brunet R and Sokolov I M, 2004 \textit{Phys. Rev. E} \textbf{70} 066102

\bibitem{Menche:2010}
Menche J,  Valleriani A and Lipowsky R, 2010 \textit{Phys. Rev. E} \textbf{81} 046103

\bibitem{Weber:2007}
Weber S and Porto P, 2007 \textit{Phys. Rev. E} \textbf{76} 046111

\bibitem{Gleeson:2008}
Gleeson J P, 2008 \textit{Phys. Rev. E} \textbf{77} 046117

\bibitem{Raschke:2010}
Raschke M,  Schl\"apfer, M  and Nibali R, 2010 \textit{Phys. Rev. E} \textbf{82} 037102

\bibitem{Sklar:1959}
Sklar A, 1959 \textit{Publ. Inst. Statist. Univ. Paris} \textbf{8} 229

\bibitem{Mari:2001}
Mari D D and Kotz S, 2001 \textit{Correlation and Dependence} (London: Imperial College Press)

\bibitem{Embrechts:2005}
McNeil A J, Frey R and Embrechts P, 2005 \textit{Quantitative Risk Management: Concepts, Techniques, Tools} (Princeton: Princeton Univ. Press)

\bibitem{Nelsen:2006}
Nelsen R B, 2006 \textit{An Introduction to Copulas} (New York: Springer)

\bibitem{Balakrishnan:2009}
Balakrishnan N  and Lai C D, 2009 \textit{Continuous Bivariate Distributions} (New York: Springer).

\bibitem{Gumbel:1960}
Gumbel E J, 1960 \textit{J. Amer. Statist. Assoc.} \textbf{55} 698

\bibitem{Catanzaro:2005}
Catanzaro M,  Bogu\~n\'a M and Pastor-Satorras R, 2005  \textit{Phys. Rev. E} \textbf{71} 027103

\end{thebibliography}
\end{document}